\documentclass[a4paper,letters,usenatbib]{mnras}

\pdfoutput=1 
\usepackage[T1]{fontenc}
\usepackage{ae,aecompl}

\usepackage{graphicx}	
\usepackage{amsmath}	
\usepackage{amssymb}	
\usepackage{txfonts}
\usepackage{hyperref}
\usepackage{natbib}
\bibpunct{(}{)}{;}{a}{}{,} 
\newcommand{\hvcs}{HVCs}
\newcommand{\ism}{ISM}
\newcommand{\rcloud}{R_{\rm cloud}}
\newcommand{\mcloud}{M_{\rm cloud}}
\newcommand{\tism}{T_{\rm ISM}}
\newcommand{\nism}{n_{\rm ISM}}
\newcommand{\mmax}{M_{\rm BE}}
\newcommand{\mjeans}{M_{\rm J}}
\newcommand{\lel}{\lambda_{\rm e}}

\title[Modeling cold gas clouds in a hot plasma]{Insights into the physics when modeling cold gas clouds in a hot plasma}

\author[B. Sander \& G. Hensler]{
Bastian Sander,$^{1,2}$\thanks{E-mail: bastian.sander@univie.ac.at, gerhard.hensler@univie.ac.at}
Gerhard Hensler,$^{1}$\footnotemark[1]
\\
$^{1}$Institut f\"ur Astrophysik, Universit\"at Wien, T\"urkenschanzstra\ss e 17, A-1180 Vienna, Austria\\
$^{2}$Fraunhofer-Institut f\"ur Fabrikbetrieb und -automatisierung IFF, Sandtorstra\ss e 22, 39106 Magdeburg, Germany
}

\date{Accepted XXX. Received YYY; in original form ZZZ}

\pubyear{2019}

\begin{document}
\label{firstpage}
\pagerange{\pageref{firstpage}--\pageref{lastpage}}
\maketitle

\begin{abstract}
This paper aims at studying the reliability of a few frequently raised but not proven arguments for the modeling of cold gas clouds embedded in or moving through a hot plasma and at sensitizing modelers to a more careful consideration of unavoidable acting physical processes and their relevance. At first, by numerical simulations we demonstrate the growing effect of self-gravity on interstellar clouds and, by this, moreover argue against their initial setup as homogeneous. We apply the adaptive-mesh refinement code {\sc Flash} with extensions to metal-dependent radiative cooling and external heating of the gas, self-gravity, mass diffusion, and semi-analytic dissociation of molecules and ionization of atoms. We show that the criterion of Jeans mass or Bonnor-Ebert mass, respectively, provides only a sufficient but not a necessary condition for self-gravity to be effective, because even low-mass clouds are affected on reasonable dynamical timescales. The second part of this paper is dedicated to analytically study the reduction of heat conduction by a magnetic dipole field.We demonstrate that in this configuration, the effective heat flow, i.e. integrated over the cloud surface, is suppressed by only $32$ per cent by magnetic fields in energy equipartition and still insignificantly for even higher field strengths.
\end{abstract}




\section{Introduction}\label{sec:intro}

Today, computer simulations are a well-established approach for describing complex astrophysical systems and supplement observational and theoretical astrophysics by a third branch known as computational astrophysics. Simulations are essential for understanding systems whose evolution cannot be adequately described via analytical means. Here, galaxies are particularly well-suited objects of study, since they contain many interacting constituents. These include already existing stars and the interstellar medium (\ism), which comprises various co-existing physical states of gas (so-called `phases') that can even lead to star formation. To add to the complexity, the \ism{} is determined by energetic processes and evolves dynamically in each of its different phases. Thus, for example, cold, interstellar (IS) gas clouds may be engulfed by hot gas - a process attracting particular research interest in the context of galactic matter cycles. When IS clouds are numerically investigated, the relevant physical processes must be modeled so as to jibe with observations, and the initial conditions must be established properly. In the literature on simulating IS clouds without star formation, there are two striking simplifications in setting up cold gas clouds in a multiphase \ism:
\begin{enumerate}
\item Self-gravity is neglected if the cloud mass is below its Jeans mass \citep[e.g.][]{09kwaksheltonraley} or the virial ratio in the cloud is much greater than one \citep[e.g.][]{17armillottaetal}. As a consequence, IS clouds are considered single-phase, homogeneous, and isothermal. The assumption is that self-gravity has a negligible effect on the evolution of low-mass clouds.
\item Heat conduction is neglected if a magnetic field in the cloud is considered \citep[e.g.][]{04mallerbullock,06esquiveletal,09kwaksheltonraley}. The assumption is that magnetic fields substantially hamper thermal conduction from the hot phase towards the cold cloud.
\end{enumerate}
But these two simplifications lack a proof, or even an estimate from physical principles. In the first part of this paper (\S~\ref{sec:initial}) we investigate whether self-gravity is a necessary initial condition when simulating the evolution of a low-mass cloud, i.e. one whose cloud mass is less than its Jeans mass. We argue that the Jeans criterion (or, more precisely, the Bonnor-Ebert criterion), however, is only a sufficient but not a necessary condition, because its derivation only describes the increase of perturbations. We are encouraged by studies such as that of \citet[][]{92habeohta}, who simulated supersonic, head-on collisions between non-identical clouds. They found that a gravitational instability in the cloud can be triggered by such collisions even if the initial cloud mass is well below its Jeans mass. Furthermore, observations of real IS clouds reveal a core-halo density structure with radial profile $\propto r^{-2}$. This is a clear sign of the role played by self-gravity \citep[][]{14kaminskietal,16wyrowskietal}, which was already deduced by \citet[][]{81larson} for virialized clouds. The observed head-tail density structure of the majority of high-velocity clouds \citep[\hvcs,][]{00bruensetal,06benbekhtietal,13forstaveleysmithmccluregriffiths} teaches us that Rayleigh-Taylor instability is suppressed due to self-gravity \citep[][]{93murrayetal,02henslervieser}. We show that even a homogeneous low-mass cloud develops a radial density profile when self-gravity is considered initially.

In the second part of the paper (\S~\ref{sec:dipolefield}) we analytically address thermal conduction by electrons from a hot gas phase towards an embedded IS cloud that exhibits a strong magnetic dipole field, which is in equipartition with the energy density of the cloud. Within the model of the three-phase \ism{} \citep[][]{77mckeeostriker} cold gas clouds represent the cold neutral phase enclosed in a warm, slightly ionized phase with both being embedded in a hot phase. Since the phases are in mutual pressure equilibrium they must have different temperatures thus thermal conduction towards the cold clouds is a natural consequence. The hot phase is very dilute hence the mean free path of electrons \citep[see e.g.][]{92shu} is of the order of $\sim 10~$pc, adopting $T\sim 10^6~$K and $n\sim 10^{-3}~$cm$^{-3}$. We go into the frequent argument that magnetic fields reduce the mean free path of electrons, by this, diminishing the effect of heat conducted by electrons to zero. We demonstrate that the effective heat flow, i.e. integrated over the cloud surface, is not suppressed substantially.

The studies of self-gravity and heat conduction presented in this paper are necessary to prepare simulations of cold gas clouds embedded in a hot, tenuous plasma. We investigate both processes separated from each other in order to evaluate their respective effect on the evolution of cold gas clouds. A prime example of particular interest in astronomy are \hvcs, which move through the hot halo of our Milky Way and contribute significantly to the galactic gas-accretion rate \citep[][]{16richter}. \hvcs{} have to be numerically investigated as it is not possible to disentangle the complex picture of the galactic matter cycle from observations alone.

\section{Self-gravity in low-mass clouds}\label{sec:initial}
\subsection{Initial setups and simulations}\label{subsec:initial}

We carry out three-dimensional, hydrodynamical simulations by using the publicly available {\sc Flash} code \citep[][]{00fryxelletal}, version 3.2\footnote{see \href{http://flash.uchicago.edu/site/flashcode/}{http://flash.uchicago.edu/site/flashcode/}} \citep[see the reference paper by][]{09dubeyetal}. It is an Eulerian hydrodynamics code, which integrates the (inviscid) Euler equations on a Cartesian, adaptively refined grid by means of the piecewise parabolic method \citep[PPM,][]{84collelawoodward,84woodwardcolella}. Our chosen computational domain is a cube with side length of $260~$pc and a finest numerical resolution of $\Delta x=1.0~$pc.

To study the effect of self-gravity on low-mass IS clouds in an isolated manner we put the clouds at rest with respect to the ambient gas. By that, internal cloud dynamics is not triggered by external forces. We focus on three models as indicated in Table~\ref{table:selfgrav-1}. First, a fiducial cloud without both self-gravity and plasma cooling is simulated to check the numerics of the code (model H0). Second and third: one cloud without (model H1) and another with self-gravity (model H2) are computed. We use {\sc Flash}'s Multigrid solver to integrate Poisson's equation. This solver is adequate for handling arbitrary density distributions.

For comparison reasons, each cloud is initially homogeneous, isothermal, and isobaric and their masses and radii are the same: $\mcloud = 6.4\times 10^4$~M$_\odot$ and $\rcloud = 41~$pc, respectively. The ambient gas has a temperature and density of $\tism = 5.6\times 10^6~$K and $\nism = 7\times 10^{-4}$~cm$^{-3}$, respectively. Temperature and density are chosen in such a way, that the model clouds and the ambient hot phase are in pressure equilibrium at $\sim 4,000~$K~cm$^{-3}$, which is in the favoured range of $10^3$ to $10^4~$K~cm$^{-3}$ for a three-phase \ism. According to \citet[][]{17tumlinsonpeepleswerk} a value of $\tism = 5.6\times 10^6~$K is consistent with observations of the Milky Way circumgalactic medium (CGM) ranging from $10^6$ to $10^7~$K. The density used in the models agrees with current measurements of the halo \citep[$10^{-5}$ to $\sim 0.002~$cm$^{-3}$,][and references therein]{16blandhawthorngerhard}. Internally, the clouds are in thermal equilibrium. Decreasing the temperature to the widely accepted value of $\tism \sim 2\times 10^6~$K \citep[][]{15millerbregman} would lead to an external pressure of $\sim 1,400~$K~cm$^{-3}$. In order to get a model cloud at the same cloud temperature and with the same mass, the cloud density, $n_{\rm old}$, must be lowered by the same factor at which $\tism$ decreases (to maintain pressure equilibrium) yielding $n_{\rm new}$. Accordingly, the cloud radius has to rise by $\left(n_{\rm old}/n_{\rm new}\right)^{1/3}$ to obtain the same cloud mass. In our case, the radius increases to $59~$pc. The decreased external pressure affects the Bonnor-Ebert mass of the cloud (cf. \S~\ref{subsec:selfgrav}), such that the ratio of cloud-to-Bonnor-Ebert mass is lower as for  $\tism = 5.6\times 10^6~$K (cf. Figure~\ref{fig:selfgrav-1}). The initial physical situation does not change and our reasoning still holds. We emphasize that we do not consider thermal conduction within the simulations.

The rates of heating and cooling of the plasma, and the degrees of dissociation and ionization are computed semi-analytically based on the local values of temperature, density, and metallicity. These local variables are calculated during runtime. Depending on temperature we have applied three different cooling laws from literature: for $T<900~$K we use the cooling curve from \citet{85falgaronepuget}, which considers molecular line cooling. For $900<T/{\rm K}<10^4$ we use the cooling function from \citet{72dalgarnomccray}, and for temperatures above $10^4~$K the plasma cools according to the rates in \citet{89boehringerhensler}. The plasma is heated owing to the photoelectric effect on dust particles \citep{06weingartnerdrainebarr}, ionization by UV radiation, by X-rays, and by cosmic rays \citep[][]{03wolfireetal}, thermalization of turbulent motions, and condensation of molecular hydrogen on dust particles \citep[both in][]{10tielens}.
\begin{table}
\caption{Simulated model clouds. The respective physical processes are either considered ($+$) or disregarded ($-$).}
\label{table:selfgrav-1}      
\centering          
\begin{tabular}{l c c}  
\hline
Model & self-gravity & heating \& cooling \\ 
\hline
H0 & $-$ & $-$ \\
H1 & $-$ & $+$ \\
H2 & $+$ & $+$ \\
\hline                  
\end{tabular}
\end{table}
\subsection{Simulation results}\label{subsec:selfgrav}

\begin{figure}
\centering
\includegraphics[width=\linewidth]{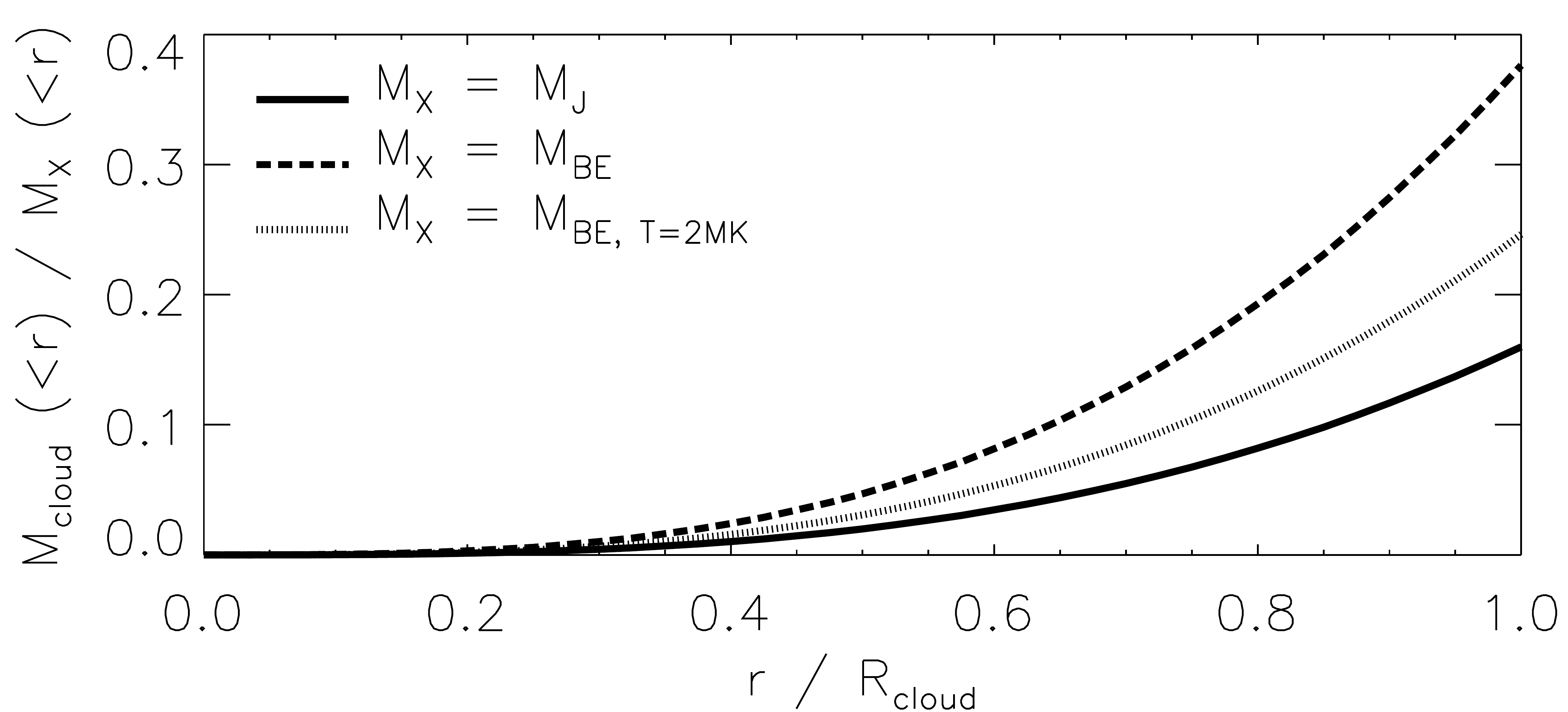}
\caption{Radial distribution of cloud-to-Jeans-mass ratio (\emph{solid line}) and cloud-to-Bonnor-Ebert mass ratio (\emph{dashed line} for $\tism =5.6\times 10^6~$K, \emph{dotted line} for $\tism =2\times 10^6~$K). The x-axis is normalized to the respective radius of each cloud.}
\label{fig:selfgrav-1}
\end{figure}

\begin{figure}
\centering
\includegraphics[width=\linewidth]{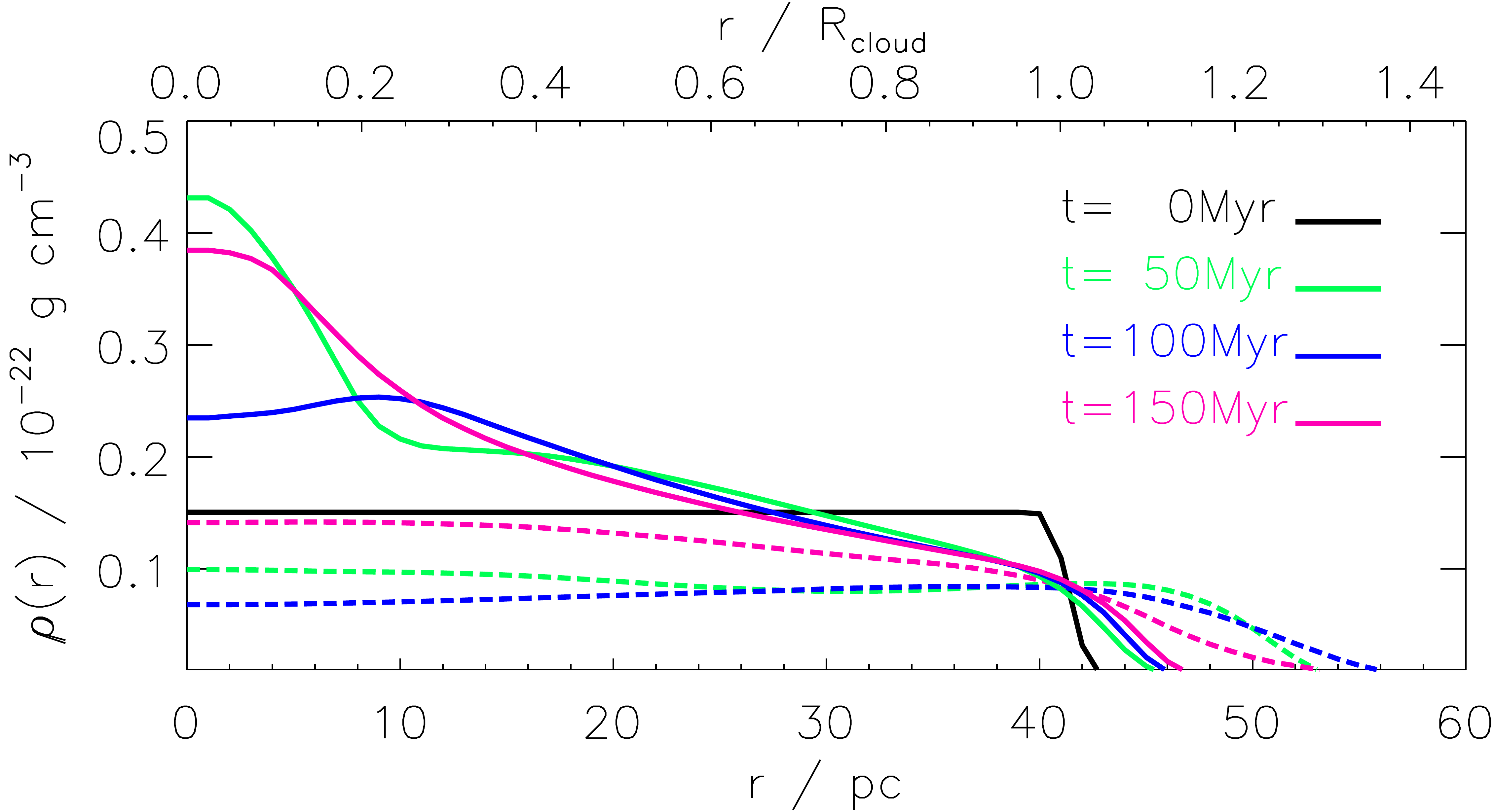}
\caption{Evolution of the angle-averaged radial density profile in model H1 (\emph{dashed lines}) and in model H2 (\emph{solid lines}). The \emph{black line} shows the initial density profile for both model clouds.}
\label{fig:selfgrav-2}
\end{figure}

\begin{figure}
\centering
\includegraphics[width=\linewidth]{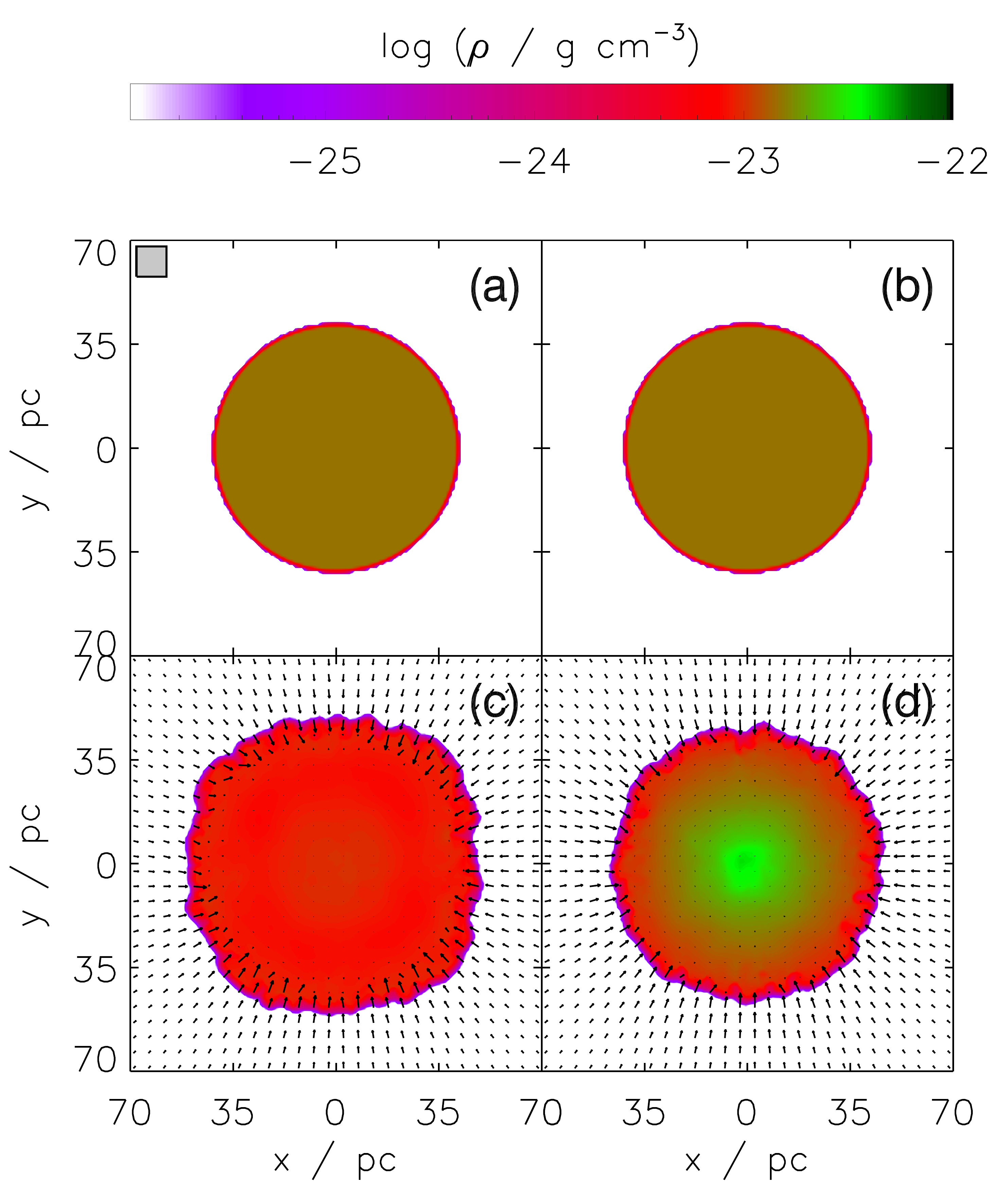}
\caption{Evolution of density in the model clouds: \emph{(a)} at initial stage, \emph{(b)} fiducial model H0, \emph{(c)} model H1, and \emph{(d)} model H2 each after $150~$Myr. \emph{Black arrows} visualize the velocity field of gas. The \emph{grey square} in the upper left corner is spanned by $10\times 10$ cells of finest resolution ($\Delta x=1.0~$pc).}
\label{fig:selfgrav-3}
\end{figure}

From linear perturbation analysis it is well-known that self-gravity in clouds is important if at any radius $r\leq \rcloud$ the \emph{Jeans mass} \citep[e.g.][]{87binneytremaine}
\begin{equation}
\mjeans=\frac{\pi}{6}\left(\frac{\pi k\bar{T}}{G\mu}\right)^{3/2}\bar{\varrho}^{-1/2}\label{equ:selfgrav-1}
\end{equation}
is exceeded by the enclosed mass $M_r= M(\leq\! r)$. It is an indicator for a region of mass $M_r$, which is not confined by an external pressure, to be gravitationally stable ($M_r\leq \mjeans$) or instable ($M_r>\mjeans$). However, if a non-vanishing ambient pressure $P_a= P(>\!\!r)$ takes effect on a cloud with radius $r$, the upper bound for its mass in order to resist gravitational instability is provided by its \emph{Bonnor-Ebert mass} \citep{55ebert,56bonnor}
\begin{equation}
\mmax=1.18\left(\frac{k\bar{T}}{\mu}\right)^2G^{-3/2}P_a^{-1/2}.\label{equ:selfgrav-2}
\end{equation}
The values $\bar{T}$, $\bar{\varrho}$, $\mu$ in equations (\ref{equ:selfgrav-1}) and (\ref{equ:selfgrav-2}) denote mean values of temperature and density and the mean molecular weight, respectively, inside $r$ and $G$ is Newton's constant of gravity. Thus, for any region,
\begin{equation}
M_r/\mmax>1\label{equ:selfgrav-3}
\end{equation}
is a \emph{sufficient} condition for self-gravity to be considered in the evolution of a low-mass cloud. However, it remains to prove the \emph{necessity} of condition (\ref{equ:selfgrav-3}), namely,
\begin{equation}
\text{self-gravity is considered in a low-mass cloud}\Rightarrow M_r/\mmax>1.\label{equ:selfgrav-5}
\end{equation}
We approach this proof by means of hydrodynamical simulations (\S~\ref{subsec:initial}). Condition (\ref{equ:selfgrav-3}) is not satisfied in any of the model clouds (Figure~\ref{fig:selfgrav-1}), but models H1 and H2 differ by the consideration of self-gravity (Table~\ref{table:selfgrav-1}). Model H1 is our reference cloud: it evolves a certain radial density profile not shaped by self-gravity. Oppositely, in model H2 self-gravity is considered. We now can use the method of proof called \emph{reductio ad absurdum}: Let's assume implication (\ref{equ:selfgrav-5}) to be true. Then self-gravity would not have any effect on the evolution of the density profile in cloud H2 \emph{a priori}. Consequently, we would not see any (or only negligible) difference in the density profiles of models H1 and H2. By inspecting Figures~\ref{fig:selfgrav-2} and \ref{fig:selfgrav-3}, however, one clearly observes a strikingly different evolution of both the density distributions and the shapes of clouds H1 and H2. Both clouds condense hot ambient gas thus increase their content of thermal energy. The expansion implied cannot be compensated in model H1. Obviously, cloud H2 is able to remain compact due to self-gravity despite its Bonnor-Ebert mass is far from being reached (Figure~\ref{fig:selfgrav-1}). Therefore, a contradiction to our assumption, that implication (\ref{equ:selfgrav-5}) is true, is shown and, by that, implication (\ref{equ:selfgrav-5}) is disproved.
After $150~$Myr, cloud H1 still oscillates entirely and with increasing elongation. However, cloud H2 remains almost constant further out but only inside the central $10~$pc densities are changing. The main differences are:
\begin{enumerate}
\item The central density increases in cloud H2 by a factor $\gtrsim 3$ within $150~$Myr. The cloud remains spherical and compact.
\item Cloud H1 remains homogeneous over $150~$Myr of evolution and expands significantly.
\end{enumerate}
We conclude, that self-gravity has a non-negligible effect on the evolution of even low-mass clouds and must in general be included to evolutionary models. Moreover, a homogeneous distribution of density does not provide a realistic condition in the presence of self-gravity.

In plot (b) in Figure~\ref{fig:selfgrav-3} it is seen that the fiducial model H0 presents a stationary solution as expected for a homogeneous cloud, which does not consider any processes for energy exchange with the surrounding gas (i.e. plasma cooling and heating), where no gravitational forces are present, and which is in pressure equilibrium with its surroundings.

\section{Heat conduction in a magnetic dipole field}\label{sec:dipolefield}

Since cool IS clouds are reasonably embedded in a warmer (up to hot) IS gas, thermal conduction determined by the temperature gradient is unavoidable. The path of charged particles like, e.g. electrons is hampered by magnetic fields. \citet{07vieserhensler1} have calculated that the mean free path of an electron perpendicular to a magnetic field in pressure equilibrium with the gas is reduced by a factor of $10^7$ according to the Larmor radius with respect to collisions with neutral hydrogen atoms. This might lead to the general and frequently expressed but nowhere documented argument, that thermal conduction cannot play a role for the energy budget of IS clouds.

A magnetic dipole field in spherical coordinates reads
\begin{equation}
\vec B=(B_r,B_\theta,B_\varphi)=\frac{m}{4\pi r^3}(2\cos\theta,\sin\theta,0),\label{eq:dipolefield-1}
\end{equation}
with radius $r$, elevation angle $\theta$, azimuthal angle $\varphi$, and magnetic moment $m$ (i.e. strength of the dipole). In presence of large temperature gradients the heat flow becomes saturated and can be written as \citep[][]{77cowiemckee}
\begin{equation}
q_{\rm sat}=0.4n_{\rm e}\left(\frac{2kT_{\rm e}}{\pi m_{\rm e}}\right)^{1/2}kT_{\rm e}\approx 5\Phi_s\varrho c_s^3,\label{eq:dipolefield-2}
\end{equation}
where $v_{\rm e}\equiv|\vec v_{\rm e}|=\left[2kT_{\rm e}/(\pi m_{\rm e})\right]^{1/2}$ is the thermal velocity of electrons, $c_s$ is the speed of sound, $\varrho$ is the density, and $\Phi_s$ covers both uncertainties owing to the treatment of flux-limited diffusion and the impact of potential magnetic fields. The saturation of heat flow being inherent in representation (\ref{eq:dipolefield-2}) accounts for a replenishment of the reservoir of heat-conducting electrons within the travel time of sound waves, i.e. there is a maximum amount of thermal energy that can be transported by electrons moving at $v_{\rm e}$. So, if all available electrons already contribute to the heat flow, its amplitude cannot increase anymore even if the temperature gradient increases \citep[see discussion in][]{06tilleybalsarahowk}.

Assuming a cloud of temperature $T_{\rm cl}$ embedded in an external medium of temperature $T_{\rm e}>T_{\rm cl}$, the temperature gradient across the cloud surface points radially inwards and hence the undisturbed flow lacks any angular component. So,
\begin{equation}
\vec q_{\rm sat}\propto \vec v_{\rm e}\parallel \vec r.\label{eq:dipolefield-3}
\end{equation}
We may thus write for the electron velocity in spherical coordinates
\begin{equation}
\vec v_{\rm e}=(v_r,0,0)\label{eq:dipolefield-4}
\end{equation}
if no local forces are present, which may be due to, e.g. turbulence or density inhomogeneities. Assume the cloud to have a magnetic dipole field (\ref{eq:dipolefield-1}). The angle $\alpha$ between magnetic field (\ref{eq:dipolefield-1}) and heat flow (\ref{eq:dipolefield-3}) is given by the dot product
\begin{equation}
\cos\alpha=\frac{\vec B\cdot\vec q_{\rm sat}}{|\vec B|\cdot|\vec q_{\rm sat}|}\propto\frac{\vec B\cdot\vec v_{\rm e}}{|\vec B|\cdot|\vec v_{\rm e}|}=\left(1+\frac{1}{4}\tan^2\theta\right)^{-1/2}.\label{eq:dipolefield-5}
\end{equation}
The radial velocity component (\ref{eq:dipolefield-4}) can be split into a component parallel, $v_\parallel$, and a component perpendicular, $v_{\rm \perp}$, to the magnetic field lines. So,
\begin{eqnarray}
v_\parallel&=&v_r\cos\alpha \\ \label{eq:dipolefield-6}
v_{\rm \perp}&=&\sqrt{v_r^2-v_\parallel^2}.\label{eq:dipolefield-7}
\end{eqnarray}
By that, the heat flow (\ref{eq:dipolefield-2}) can be split in the same way. That is,
\begin{eqnarray}
q_\parallel&=&0.1n_{\rm e}v_\parallel\frac{3}{2}kT_{\rm e}\\ \label{eq:8}
q_{\rm \perp}&=&0.1n_{\rm e}v_{\rm \perp}\frac{3}{2}kT_{\rm e},\label{eq:9}
\end{eqnarray}
such that $q_{\rm sat}=\sqrt{q_\parallel^2+q_{\rm \perp}^2}$. The Lorentz force acts on $q_{\rm \perp}$ only.

\begin{figure}
\centering
\includegraphics[width=\linewidth]{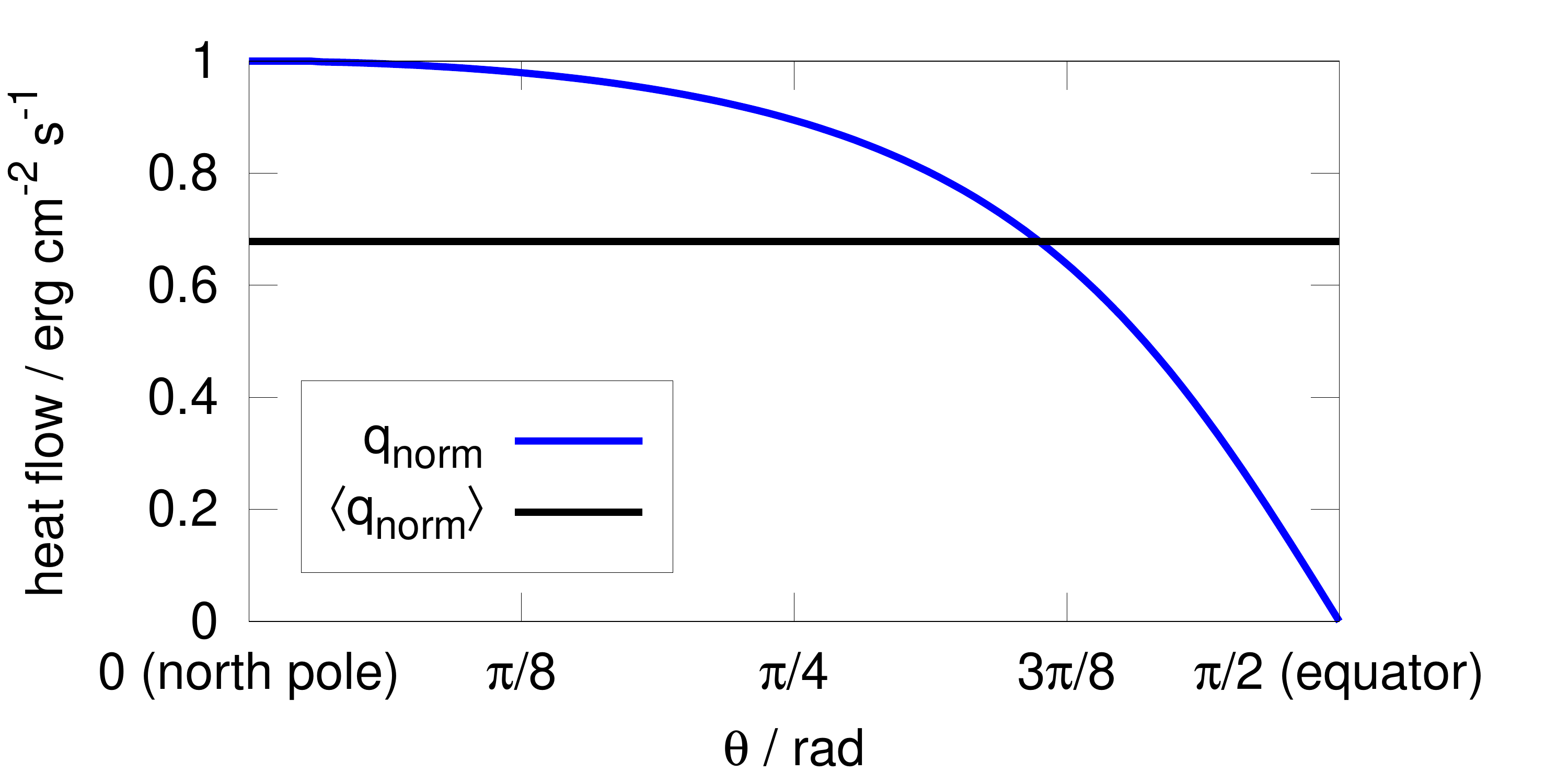}
\caption{Normalized heat flow, $q_{\rm norm}$ (\emph{blue line}), at a cloud-centric distance of $r=40~$pc. The weighted mean of normalized heat flow, $\langle q_{\rm norm}\rangle$ (\emph{black line}), is averaged over all $\theta$ ranging from 0 to $\pi/2$.
}
\label{fig:dipolefield-1}
\end{figure}

We now have to estimate the degree of suppression by the magnetic field if a specific thermal energy of electrons is given. If no magnetic field is present, the electrons travel their respective mean free path for Coulomb collisions $\lel=t_{\rm e}v_{\rm e}$, with an electron-electron equipartition time \citep{62spitzer}
\begin{equation}
t_{\rm e}=\frac{3\sqrt{m_{\rm e}}(kT_{\rm e})^{3/2}}{4\sqrt{\pi}n_{\rm e}e^4\ln(\Omega)}.\label{equ:parameter-3}
\end{equation}
If a magnetic field is present, the electrons are deflected by the Lorentz force and hence gyrate around the field lines with Larmor radius 
\begin{equation}
r_L=\frac{m_{\rm e}v_\perp}{e|\vec B|}.\label{eq:dipolefield-11}
\end{equation}
Therefore, the factor, by which $q_\perp$ is suppressed, is given by $r_L / \lambda_{\rm e}$, since $r_L$ measures the disturbing effect of the magnetic field on the straight line $\lambda_{\rm e}$. So, the heat flow, which passes the magnetic field, reads
\begin{equation}
q_{\rm eff}=\sqrt{q_\parallel^2+\left(\frac{r_L}{\lambda_{\rm e}}q_{\rm \perp}\right)^2}.\label{eq:dipolefield-13}
\end{equation}
For sufficiently weak magnetic fields $r_L\sim\lambda_{\rm e}$ thus the heat flow perpendicular to $\vec B$ is not affected, but for stronger fields $r_L$ can easily fall below $10^{-8}\lambda_{\rm e}$ \citep[see discussion in][]{07vieserhensler1}. The normalized heat flow $q_{\rm norm}\equiv q_{\rm eff}/q_{\rm sat}$ measures the fraction of the flow (\ref{eq:dipolefield-2}) that is able to pass the magnetic field. The magnetic moment is chosen to be $m\approx 3\times 10^{51}$ erg G$^{-1}$, such that $|\vec B|\sim 5~\mu$G at cloud surface ($r=40~$pc) at equator ($\theta=\pi/2$), which corresponds to equipartition with the energy density of the cloud. In Figure~\ref{fig:dipolefield-1} a clear dependence of $q_{\rm norm}$ on the elevation angle $\theta$ is visible. At equator a value of $q_{\rm norm}\sim 10^{-9}$ is established.

Averaging $q_{\rm norm}$ over all $\theta$ yields the weighted mean
\begin{equation}
\langle q_{\rm norm}\rangle=\frac{\int\limits_{\theta=0}^{\pi/2}\int\limits_{\varphi=0}^{2\pi}\sin\theta'q_{\rm norm}(\theta'){\rm d}\theta'{\rm d}\varphi'}{\int\limits_{\theta=0}^{\pi/2}\int\limits_{\varphi=0}^{2\pi}{\rm d}\theta'{\rm d}\varphi'}=\frac{2}{\pi}\int\limits_{\theta=0}^{\pi/2}\sin\theta'q_{\rm norm}(\theta'){\rm d}\theta'.\label{eq:dipolefield-15}
\end{equation}
As can be seen in Figure~\ref{fig:dipolefield-1}, $\langle q_{\rm norm}\rangle$ is suppressed to $68$ per cent only. More extensive, but rather academic numerical studies on the impact of different magnetic field configurations on heat flow have been performed by \citet{12lifrankblackman}, taking into account the dynamic evolution of the magnetic field. Their heat-transfer efficiency $\zeta$ has the same meaning like our $q_{\rm norm}$. What they observe is a hampered conduction if the magnetic field is more tangled, i.e. if the contributions from field lines perpendicular to the heat flow are  increased (their situation for $\zeta\to 0$). This corresponds to $q_{\rm norm}\left(\theta=\pi/2\right)\sim 10^{-9}$ (Figure~\ref{fig:dipolefield-1}). However, they figured out that if the global magnetic field dominates the local tangled field (in their terminology: $\zeta\to 1$), the impediment imposed by anisotropic heat conduction in the local tangled field is negligible. This result conforms with $q_{\rm norm}(\theta=0)=1$. The findings of \citet{12lifrankblackman} hence substantiate our results even though they have not averaged all particular heat flows for $0\leq\zeta\leq 1$ to get an effective heat flow.

\section{Summary and conclusions}\label{sec:sumcon}

The aim of our exploration is twofold. In the first part (\S~\ref{sec:initial}) we present three-dimensional, hydrodynamical simulations of three models of resting low-mass clouds (Figure~\ref{fig:selfgrav-1}), which are initially homogeneous (\S~\ref{subsec:initial}). Our main purpose of these studies is to analyze whether self-gravity plays a substantial role in the evolution of cold gas clouds with masses well below their Jeans mass. After $150~$Myr of evolution the morphologies of our model clouds H1 and H2 are strikingly different (Figure~\ref{fig:selfgrav-3}). An initially homogeneous cloud without self-gravity (model H1) remains homogeneous and is shaped irregularly. The evolutionary process is totally different in the case of self-gravity (model H2), where the cloud remains compact, roundish, condenses centrally, and evolves a radial density gradient (Figure~\ref{fig:selfgrav-2}). Hence we conclude that self-gravity has a major impact on even gravitationally stable clouds ($M<\mmax$), which disproves the necessity of condition (\ref{equ:selfgrav-3}). Also a homogeneous density distribution is not a realistic initial condition if self-gravity is accounted for.

We further calculated analytically the effective heat flow of electrons passing a magnetic dipole field, which contributes major to a multipole field. It turned out that even in equipartition the heat flow, which is integrated over the cloud's surface, is reduced to $68$ per cent only (Figure~\ref{fig:dipolefield-1}). The evaluation of the degree of suppression of heat conduction becomes even more complicated for more complex configurations of magnetic fields \citep[e.g. for tangled fields due to hydrodynamic instabilities or turbulence,][]{08guoohruszkowski} or if the fields evolve with time. \citet{01narayanmedvedev} have shown that the effective heat flow is suppressed to only $\sim 20$ per cent if the entanglement of the magnetic field lines extends over certain spatial scales. We thus deduce that even in the presence of strong magnetic fields the heat flow must not be neglected in simulations of cool clouds embedded in a hot plasma.

The study presented is thought as a preparatory effort for cloud models with heat conduction and the evolution of \hvcs. In two forthcoming papers we show that the arguments we figured out in our analysis performed with resting clouds can be applied to fast moving low-mass clouds, e.g. \hvcs, as well.

\section*{Acknowledgements}

We gratefully acknowledge the supporting comments by an anonymous referee, which led to a substantial improvement of the clarity of this paper. This work was partially supported by Initiative College IK538001 of the University of Vienna and partially supported by the Austrian \emph{Fonds zur F\"orderung der wissenschaftlichen Forschung (FWF)} under project number AP2109721. The software used in this work was in part developed by the DOE NNSA-ASC OASCR Flash Center at the University of Chicago. The computational results presented have been achieved by using the Vienna Scientific Clusters 1 and 3 (VSC-1 and VSC-3)\footnote{see \href{http://vsc.ac.at/}{http://vsc.ac.at/}}.




\bibliographystyle{mnras}
\bibliography{references} 



%
%



\bsp	
\label{lastpage}
\end{document}